\documentclass[aps,prl,superscriptaddress,reprint]{revtex4-1}

\usepackage{amsmath}
\usepackage{amssymb}
\usepackage{graphicx}
\usepackage{amsfonts}
\usepackage{bm}
\usepackage[latin9]{inputenc}
\usepackage{mathrsfs}
\usepackage[mathscr]{eucal}
\usepackage{esint}
\usepackage{subfigure}
\usepackage{lineno}

\makeatletter

\@ifundefined{textcolor}{}
{%
 \definecolor{BLACK}{gray}{0}
 \definecolor{WHITE}{gray}{1}
 \definecolor{RED}{rgb}{1,0,0}
 \definecolor{GREEN}{rgb}{0,1,0}
 \definecolor{BLUE}{rgb}{0,0,1}
 \definecolor{CYAN}{cmyk}{1,0,0,0}
 \definecolor{MAGENTA}{cmyk}{0,1,0,0}
 \definecolor{YELLOW}{cmyk}{0,0,1,0}
}

\newcommand{\ket}[1]{\mid#1\,\rangle}
\newcommand{\bra}[1]{\langle\,#1\mid}
\newcommand{\ie}{{\it i.e.}}

\begin{document}


\hyphenation{its stands questions differs equ-ation model nature theory space state evalu-ate rotates depends another refers Since slowly thousands}



\title{Quantum Approach to Fast Protein-Folding Time}

\author{Li-Hua Lu}
\affiliation{\noindent
Zhejiang Province Key Laboratory of Quantum Technology $\&$ Device and Department of Physics, Zhejiang University, Hangzhou 310027, P.R. China}
\author{You-Quan Li}\email[email: ]{yqli@zju.edu.cn}
\affiliation{\noindent
Zhejiang Province Key Laboratory of Quantum Technology $\&$ Device and Department of Physics, Zhejiang University, Hangzhou 310027, P.R. China}
\affiliation{\noindent
Collaborative Innovation Center of Advanced Microstructure, Nanjing University, Nanjing 210008, R.R. China}

\received{\today}


\begin{abstract}

In the traditional random-conformational-search model
various hypotheses with a series of meta-stable intermediate states were often proposed
to resolve the Levinthal paradox.
Here we introduce a quantum strategy to formulate protein folding as a quantum walk on a definite graph, which provides us a general framework without making hypotheses.
Evaluating it by the mean of first passage time, we find that the folding time via our quantum approach is much shorter than the one obtained via classical random walks.
This idea is expected to evoke more insights for future studies.

\end{abstract}

\maketitle


Understanding how  proteins fold spontaneously into their native structures is both a fascinating
and fundamental problem in interdisciplinary fields involving
molecular biology, computer science, polymer physics as well as theoretical physics etc..
Since Harrington and Schellman discovered that protein-folding reactions are very fast and often reversible processes\cite{Schellman1956}, there has been progressively more investigations on protein folding
in both aspects of theory and experiment.
Levinthal~\cite{Levinthal}  noted early in 1967 that a much larger folding time is inevitable if proteins are folded by sequentially sampling of all possible conformations.
Thus the protein was assumed to fold through a series of meta-stable intermediate states
and the random conformational search does not occur in the folding process.
The questions about what are the energetics of folding and how does the denature cause unfolding
motivates one to think that
the protein folding proceeds  energetically downhill and loses conformational entropy as it goes.
Based on such a hypothesis, the free-energy landscape framework was one way to describe the protein folding~\cite{Malla,Wolynes,Passway:Jackson},
where the energy funnel landscape provided a first conjecture of how the folding begins and continues~\cite{Wolynes2012}.

As we known, there have been substantial theoretical models with different
simplifying assumptions, such as Ising-like model~\cite{Ising1,Ising2}, foldon-dependent protein folding model~\cite{Englander}, diffusion-collision model~\cite{Karplus,Sali}, and
nucleation-condensation mechanism~\cite{Thirumalai,Fersht} etc..
Theoretical models are useful for understanding the essentials of the complex self-assembly reaction of protein folding, but till now, they often rely on various hypotheses~\cite{Wolynes2005,Shakhnovich2006,Wolynes2012,Dill2012,Thirumalai2013}.
This often brings in certain difficulties in connecting analytical theory to experimental results
because some hypotheses can not be easily put into a practical  experimental measurement.
As it introduces less hypotheses in comparison to those theoretical models,
the atomistic simulations~\cite{PianaS,HenryR,Snow} were used to investigate the protein folding along with nowadays' advances in computer science.
Recently, a high-throughput protein design and characterization method
was reported that allows one to systematically examine how sequence determines the folding and stability~\cite{Rocklin2017}.
However, quantitatively achieving the folding time  and  accurately understanding how the sequence determines the protein folding remain to be  a key challenge.

Here we propose a quantum strategy to formulate protein folding as a quantum walk on a definite graph, which provides us a general scheme without artificial hypotheses.
In terms of the first-passage probability, one can calculate the folding time as the mean of the first-passage time.
The obtained folding time in terms of our quantum scheme is much shorter than the one obtained via classical random walks.
This idea is expected to open a new avenue for investigating  the protein folding theoretically,
which may motivate a necessary step toward developing technology for protein engineering and designing protein-based nanodevices~\cite{Munoz}.

\textit{Theoretical consideration:}
We describe the protein structure by the frequently adopted lattice model~\cite{GoModel,HP:Dill,LiTang1996,LiJi2004},
namely, a protein is regarded as a chain of non-own intersecting unit
(usually referring an amino-acid residue) of a given length on the two-dimensional square lattice.
For a protein with $n$ amino-acid residues,
we can calculate the total number $N_n$ of distinct lattice conformations
that distinguish various protein intermediate structures.
For instance, we have $N_4=4$ and $N_6=22$.
This provides us a set with $N_n$ objects,
which we call structure set and denote it by
$\mathscr{S}_n =\{s^{}_1, s^{}_2, \cdots, s^{}_{N_n} \}$ hereafter.

In order to study the protein folding process, we propose a concept of one-step folding.
On the basis of the lattice model, we can naturally define the one-step folding by one displacement of an amino acid in one of the lattice sites.
This makes us to establish certain connections between distinct points in the set $\mathscr{S}_n$
and to have a connection graph $\mathscr{G}_n$.
In other words, two structures are connected via one-step folding if their conformation
differs in one site only.
As a conceptual illustration,
we plotted the structure set $\mathscr{S}_4$, the connection graphs $\mathscr{G}_4$ and  $\mathscr{G}_6$ in Fig.~\ref{fig:model} (the $\mathscr{S}_6$ in Fig.~S1 in
supplementary material).
Such a graph $\mathscr{G}_n$ is described
by the so-called adjacency matrix $\mathrm{Mat}(J_{ab})$
that
characterizes a classical random walk~\cite{Kampen1997} on the graph.

\textit{Folding as a quantum walk:}
Letting $\ket{s_a}$ denote the state of a protein structure in the shape of the $a$-th lattice conformation,
we will have a quantum Hamiltonian in a $N_n$-dimensional Hilbert space,
%
\begin{equation}\label{eq:hamiltonian}
\hat{H}^{}_0 =
-\sum_{a,b} J_{a b}\ket{s^{}_a}\bra{s^{}_b},
\end{equation}
where $J_{a b}$ refers to the connection between different points in the structure set,
\ie, $J_{a b}$ is nonzero only if the $a$-th protein structure $s^{}_a$ can be transited into
the $b$-th structure $s^{}_b$ by a one-step folding.
With these physics  picture one can also investigate quantum walk~\cite{Aharonov1993,Farhi1998,WangJB2014} on the aforementioned graph.


\begin{figure}[t]
\includegraphics[width=0.42\textwidth=0.43]{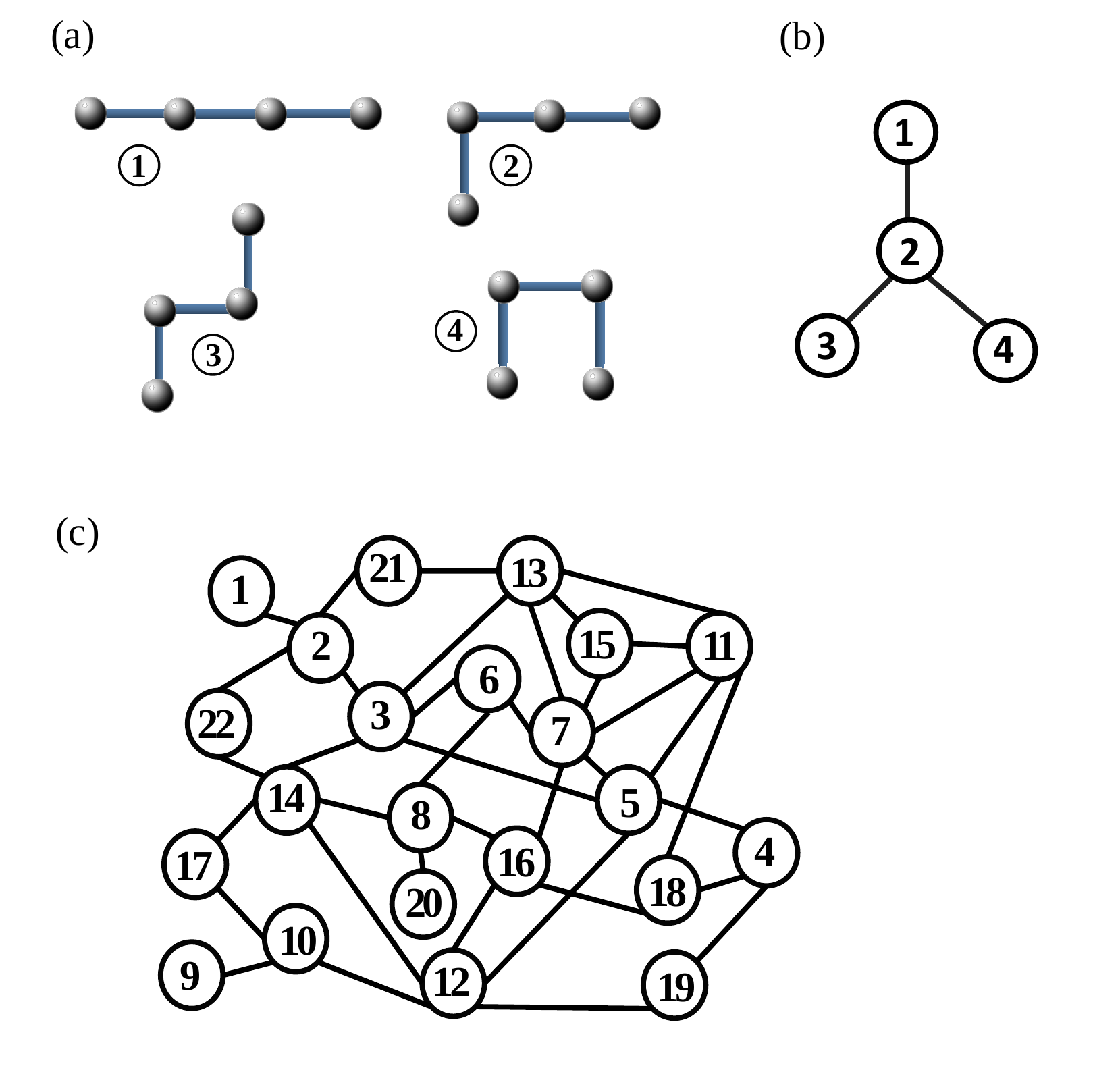}
\vspace{-2mm}
\caption{\label{fig:model}\small
Structure sets  and connection graphs.
(a) There are four distinct structures for the amino-acid chain with 4 residues,
thus the corresponding structure set $\mathscr{S}_4$ contains 4 objects.
(b) The connection graph $\mathscr{G}_4$ includes 4 vertices,
which happens to be a three-star graph.
(c) The connection graph $\mathscr{G}_6$ includes 22 vertices.}
\end{figure}



From the coarse grained point of view, the 20 amino acids are classified~\cite{HP:Dill}
as hydrophobic and hydrophilic (it is also called polar) groups according to their contact interaction.
As $H$ and $P$ represent the hydrophobic and  polar amino acids conventionally,
a sequence of $n$ amino acids can be labeled by $\vec{q}=(q^{}_1, q^{}_2, \cdots, q^{}_n)$ where $q^{}_k$
with $k=1,2, \cdots, n$ refers to either $H$ or $P$.
Thus there will be totally a set of $2^n$ possible sequences.
Let us call the entire of the whole random sequences as
the sequence set denoted by
$\mathscr{Q}_n=\{ {\small [\nu]} \mid \nu=1, 2, \cdots, 2^n\}$.
For any definite sequence-$[\nu]$ specified by a $\vec{q}$,
we can calculate the total contact energy~\cite{LiTang1996,LiJi2004}
for each structure in $\mathscr{S}_n$,
\begin{equation}\label{eq:contact}
\mathcal{E}^{[\nu]}_a =\sum_{k<l}E_{q^{}_k q^{}_l}\delta_{|\textbf{r}^a_{k}-\textbf{r}^a_{l}|,1}(1-\delta_{|k-l|,1}),
\end{equation}
where $a=1, 2, \cdots, N_n$ labels different structures,
$k$ and $l$  denote the successive labels of the amino-acid residues
in the sequence (\ie, the order in the chain),
while
$\textbf{r}^a_{k(l)}$ stands
for the coordinate position of the $k(l)$-th residue in the $a$-th structure
and $q^{}_{k(l)}$ refers to either $H$ or $P$.
Here the notation of Kronecker delta is adopted, \ie, $\delta_{\alpha,\beta}=1$ if $\alpha=\beta$ and $\delta_{\alpha ,\beta}=0$ if $\alpha\neq \beta$.
It is widely believed that the native structure of a protein possesses the lowest free energy~\cite{Anfinsen}.
This can be interpreted by the hydrophobic force that drives the protein to fold into a compact structure with as many hydrophobic residues inside as possible~\cite{HP:Dill}.
Thus the $H$-$H$ contacts are more favourite in the lattice model~\cite{HP:Dill,HP:Onuchic,HP:Shakhnovich,HP:Wolynes}, which can be characterized by choosing $E^{}_{PP}=0$, $E^{}_{HP}=-1$, and $E^{}_{HH}=-2.3$ as adopted in Ref.~\onlinecite{LiTang1996}.

With the contact energy (\ref{eq:contact}) for every structure,
the potential term can be expressed as
\begin{equation}\label{eq:potential}
V^{[\nu]} = \sum_{a} \mathcal{E}^{[\nu]}_a \ket{s^{}_a}\bra{s^{}_a}.
\end{equation}%
Thus the total Hamiltonian for a definite sequence-$[\nu]$ is given by
$ \hat{H}^{[\nu]} = \hat{H}^{}_0 + V^{[\nu]}$. 
%
Clearly, the kinetic term $ \hat{H}^{}_0$ is determined by the connection graph $\mathscr{G}_n$ merely
while the potential term $V^{[\nu]}$ defined on the structure set $\mathscr{S}_n$
is related to the concrete sequence-$[\nu]$ under consideration.
This means that we have a hierarchy of Hamiltonian
$\{ H^{[\nu]} \mid \nu =1, 2, \cdots, 2^n \}$ actually
for a theoretical study of the protein folding problem.

Note that one may obtain the same contact energy $\mathcal{E}_a$ for several different sequences.
In this case, the dynamical properties are the same although those sequences may differ.
Such a dynamical degeneracy implies a partition within  the sequence set $\mathscr{Q}_n$.
There are totally 16 possible sequences in
$\mathscr{Q}_4$ which is partitioned into three subsets, \ie,
$\mathscr{Q}_4=\{ Q_1, Q_2, Q_3 \}$,
thus there will be three situations in the discussion on the time evolution.
For $n=6$, there are totally 64 possible sequences in $\mathscr{Q}_6$
which is partitioned into 45 subsets,
\ie, $\mathscr{Q}_6 = \{ Q_1, Q_2, \cdots, Q_{45} \}$
(see Tables SI $\&$ SII in supplementary material).

\textit{Random walk with sticky vertices:}
As we known,
the continuous time classical random walk~\cite{Montroll1965} on a graph
$\mathscr{G}_n$ is described by the time evolution of the
probability distribution $p^{}_a (t)$ that obeys the master equation
\begin{equation}\label{eq:classicalRW}
\frac{\mathrm{d} }
     {\mathrm{d}t}
      p^{}_a (t) = \sum_b K^{}_{a b}\,p^{}_b (t),
\end{equation}
where  $K_{a b}=T_{ab}-\delta_{a b}$
with $T^{}_{ab}$ being the probability-transition matrix.
In the conventional classical random walk,
the probability-transition matrix is determined by the adjacency matrix of an undirected graph,
namely,
$T_{ab}=J_{ab}/\mathrm{deg}(b)$
where $\mathrm{deg}(b)=\sum_c J_{cb}$ represents the degree of vertex-$b$
in the graph $\mathscr{G}_n$.
However, we ought to reconsider the random walk
if there are some ``sticky'' vertices in the graph.
This corresponds to the case
when we take account of the contact energy $\mathcal{E}_a$
in the protein conformations.
Thus, the probability-transition matrix should be modified
so that the strength hopping into differs from that hopping out of those sticky vertices.
The modified transition matrix $\tilde{T}$ is given by
\begin{equation}\label{eq:gamma}
\tilde{T}^{}_{a b} = T^{}_{a b} - \Gamma_{a b}+\Lambda_{ab}.
\end{equation}
Here
$\Gamma_{ab}=\theta({\cal E}_{ab})\Omega_{ab} T_{ab}$,
$\Omega_{ab}={\cal E}_{ab}^2/({\cal E}_{ab}^2 + 1)$,
and
$\Lambda_{ab} = \delta_{ab}\sum_{c}\Gamma_{cb}$
where a notation ${\cal E}_{ab}={\cal E}_a -{\cal E}_b$ is adopted for simplifying the expression.
The newly added two terms in (\ref{eq:gamma}) together guarantee the probability conservation.
Therefore, in the presence of sticky vertices, one needs to solve the master equation (\ref{eq:classicalRW})
with the modified $\tilde{K}=\tilde{T}-I$ in the discussion of classical random walks.

\textit{The quantum dynamics:}
To accomplish a quantum mechanical understanding,
we take account of the energy dissipation caused by the medium in which
the folding occurs.
This is governed by the Lindblad equation~\cite{Lindblad}
\begin{equation}
\frac{\mathrm{d}}{\mathrm{d} t}\hat{\rho}
  = \frac{1}{i\hbar} [\hat{H},\hat{\rho} ] +\mathcal{L}(\hat{\rho}),
\label{eq:masterequation}
\end{equation}
where
\begin{equation}
\mathcal{L}(\hat{\rho})=
 \frac{\lambda}{2}\bigl(
 2 L\hat{\rho}\,L^{\dagger}
    -\hat{\rho}\,L^{\dagger}L
    -L^{\dagger}L \hat{\rho}
    \bigr)
\label{eq:Lindblad}
\end{equation}
reflects the effect of dissipation.
Here $L$ and $L^\dagger$ is called the Lindblad operator
which can be determined from the analyses of random walks in the presence of sticky vertices.
The aforementioned off-diagonal part $\Gamma$ in (\ref{eq:gamma}) provides this operator
{\it i.e.}, $L^{\dagger}=\sum_{a b}\Gamma_{a b}\ket{s^{}_a}\bra{s^{}_b}$.
Actually, equation (\ref{eq:Lindblad}) presents a general expression,
which becomes the traditional one in terms of Pauli matrices,
$\mathcal{L}(\hat{\rho})=
 (2\sigma^{-}\hat{\rho}\,\sigma^{+}
    -\hat{\rho}\,\sigma^{+}\sigma^{-}
    - \sigma^{+}\sigma^{-} \hat{\rho})\gamma/2$
with $\gamma=\lambda\Omega^2$
for a two level system that can be regarded as
the two-vertices graph with a sticky vertex.


\begin{figure*}[t]
\includegraphics[width=0.9\textwidth=1.56]{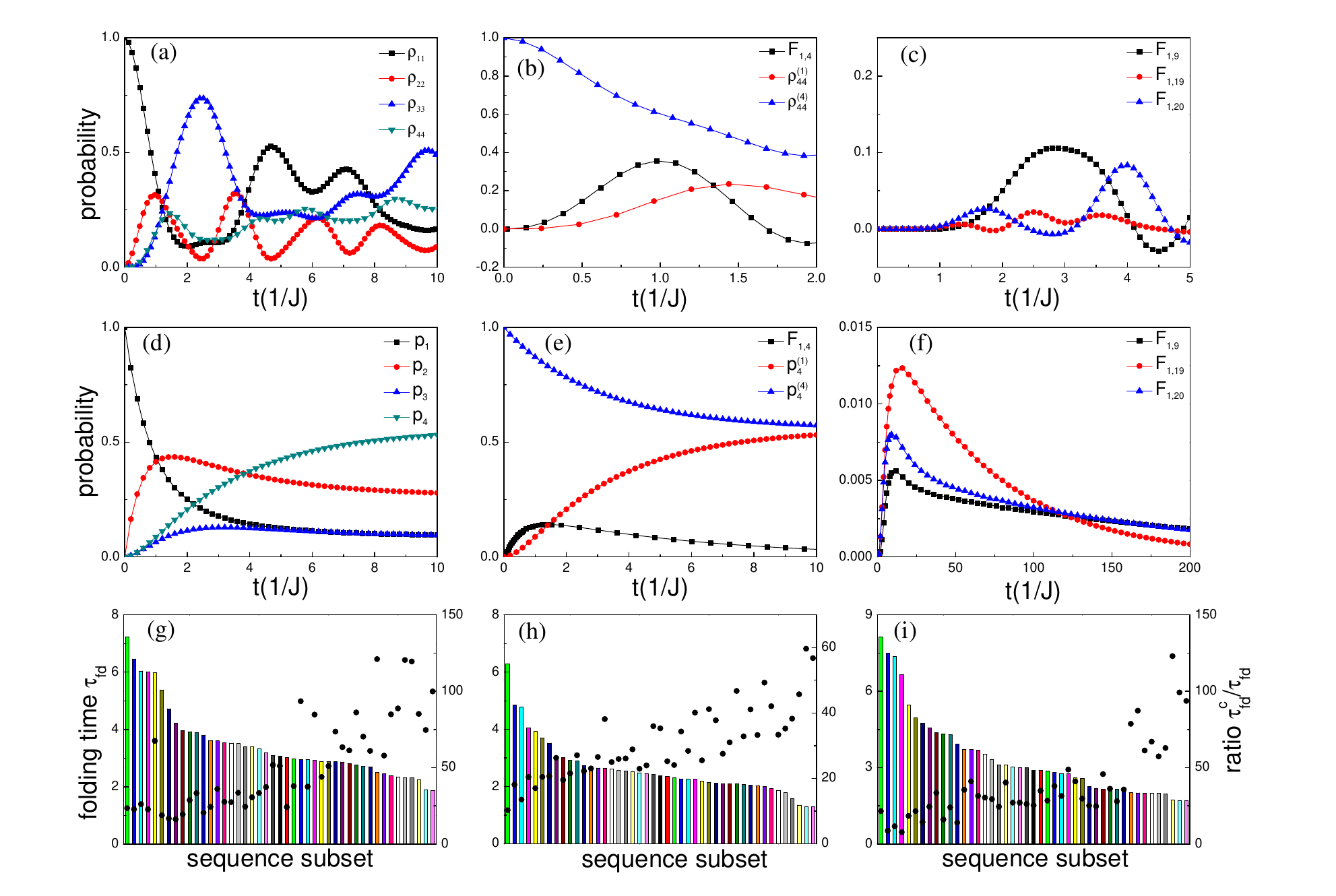}
  \vspace{-1mm}
\caption{\label{fig:collections}\small
Illustrations for the folding dynamics and the comparison of the folding times.
(a) The time evolution of the diagonal elements of density matrix.
(b) The quantum folding process for the sequence subset $Q_3$ with $n=4$.
It is $\rho^{(1)}_{44}$ together with $\rho^{(4)}_{44}$ that determines the first-passage probability $F^{}_{1,4}$ which reaches zero when $t=1.7$ and becomes negative afterwards.
(c) The solved first-passage probabilities concerning to the quantum folding process on the graph $\mathscr{G}_6$.
$F_{1,9}$, $F_{1,19}$ and $F_{1,20}$ are the data for sequence-$[\mathit{37}]$
when taking structures-9, -19 and -20 as the target states respectively.
Here
$\tau^{}_0= 4.12$, $\tau^{}_0= 1.70$, and $\tau^{}_0=2.44$ correspondingly.
(d) The time evolution $p_1(t)$, $p_2(t)$, $p_3(t)$ and $p_4(t)$ of classical random walk on $\mathscr{G}_4$.
(e) The classical folding process for subset $Q_3$.
The solved first-passage probability is positive at finite time and approaches to zero when
$t$ goes to infinite.
(f) The classical folding process on $\mathscr{G}_6$ for sequence-$[\mathit{37}]$
as a comparison to the quantum case.
(g) to (i) Quantum folding time $\tau^{}_\mathrm{fd}$ and the ratios of classical folding time $\tau^\mathrm{c}_\mathrm{fd}$ to quantum ones.
The former is plotted in terms of histogram which is scaled by the left vertical axis,
the latter is plotted by black dots which is scaled by the right vertical axis.
Correspondingly, they are the data respectively  with the most compact structures $s_9$ (g), $s_{19}$ (h) and  $s_{20}$ (i) as the folding targets.
}
\end{figure*}


We solve the density matrix $\hat \rho(t)$
from Eq.~(\ref{eq:masterequation})
with the initial condition
$\hat{\rho}(0)=\mid\! s_1 \rangle \langle s_1\! \mid$.
Here $\mid s_1 \rangle$ refers to the completely unfolded straight-line structure.
To illustrate our theory intuitively, we start from the simplest model of $n=4$
where the protein-folding problem
becomes a task to investigate the quantum walk on the graph $\mathscr{G}^{}_4$.
We solve the  $\hat{\rho}(t)$ numerically for the three situations $Q_1$, $Q_2$ and $Q_3$ respectively.
In the calculation, we set $\hbar$ and $J$ to be unity
and take the time step as $\Delta t=0.02$.
For the initial condition:
 $\rho^{}_{1 1}=1$ and the other matrix elements vanish when $t=0$,
we solve Eq.~(\ref{eq:masterequation})
by means of Runge-Kutta method and obtain the magnitude of $\rho_{ab}^{(1)}(t)$ at any later time,
$t=j*\Delta t$  with $j=1,2,\cdots$.
We plot the time dependence of the diagonal elements of the solved
density matrix for the $Q_3$ case in Fig.~\ref{fig:collections}(a) and the other cases
in the supplementary material Fig.~S3. 
Likewise, we solve the density matrix for another initial condition
$\rho^{}_{44}(0)=1$ again so that the first-passage probability can be determined
later on.
The population of the most compact structure $\ket{s_c}$  is evaluated by the
diagonal element $\rho^{}_{cc}(t)$.
For instance, $c=4$ in $\mathscr{S}_4$,
and $c=9, 19,$ \textrm{and}  $20$ in $\mathscr{S}_6$.
We can see that the probability of the state
referring to the most compact structure $\ket{s_4}$ increases
much more rapidly in the quantum folding process Fig.~\ref{fig:collections}(a) than in the classical process Fig.~\ref{fig:collections}(d).
Toward a genuine understanding, we further study
the quantum walk on the graph $\mathscr{G}^{}_6$ by solving the density matrix numerically one by one
for the aforementioned forty-five situations.

\textit{The folding time:}
Now we are in the position to define the protein folding time
which can be formulated with the help of the concept of the mean first-passage
time~\cite{FirstPassageTime1969,FirstPassageTime2001,FirstPassageTime2004,FirstPassageTime2007,FirstPassageTime2016}.
The mean first-passage time from a starting state $\ket{s^{}_a}$ to a target state $\ket{s^{}_b}$ is given by
$
\int_{t=0}^{\tau^{}_0} t F_{a,b}(t)\mathrm{d}t\,/\int_{t=0}^{\tau^{}_0} F_{a,b}(t)\mathrm{d}t
$
where $\tau^{}_0$ represents the time period when the first-passage probability vanishes $F_{a,b}(\tau^{}_0)=0$ which really occurs for the aforementioned quantum walk.
For example in Fig.~\ref{fig:collections}(b), the solved first-passage probability $F_{1,4}(t)$ becomes negative after
$t=1.7$.
The first-passage probability $F^{}_{a,b}(t)$ from a state $\ket{s^{}_a}$ to another state $\ket{s^{}_b}$ after $t$ time obeys the known convolution relation
\begin{equation}\label{eq:FPTprobability}
P^{}_{a,b}(t)=
\int^{t}_{0} F^{}_{a,b}(t')P^{}_{b,b}(t-t')\mathrm{d}t'.
\end{equation}
Here $P^{}_{a,b}(t)$ denotes the probability of a state being the basis state $\ket{b}$ at time $t$ if starting from the state $\ket{a}$ at initial time $t=0$.
Quantum mechanically, it is evaluated by the diagonal elements of the density matrix,
{\it i.e.},
$P^{}_{a,b}(t)=\rho^{(a)}_{b b}(t)$
where
$\rho^{(a)}_{b b}(t)=\bra{b}\hat{\rho}^{(a)}(t)\ket{b}$
is solved from Eq.~(\ref{eq:masterequation})
with the initial condition $\hat{\rho}(0)=\ket{a}\bra{a}$,
while $P^{}_{b,b}(t)=\rho^{(b)}_{b b}(t)$ is solved with another initial condition $\hat{\rho}(0)=\ket{b}\bra{b}$.
Here the superscripts are introduced to distinguish the solution from different initial conditions.
In the classical case, $P^{}_{a,b}$ and $P^{}_{b,b}$ refer to the $p^{}_{b}(t)$ solved from Eq.~(\ref{eq:classicalRW}), respectively, with initial conditions $p^{}_c (0) =\delta^{}_{ac}$
and $p^{}_c (0) =\delta^{}_{bc}$.

As protein folding is the process that proteins achieve their native structure,
the folding time is the case that the starting state is chosen as $\ket{s^{}_1}$ and the target states are the most compact states.
For example, they are $\ket{s^{}_9}$, $\ket{s^{}_{19}}$ or $\ket{s^{}_{20}}$
for $n=6$.
The formula for the calculation of the folding time is thus given by
\begin{equation}\label{eq:fdtime}
\tau^{}_\mathrm{fd}=
 \frac{\int_{0}^{\tau^{}_0 } t F^{}_{1,c}(t)\mathrm{d}t}
      {\int_{0}^{\tau^{}_0 }  F^{}_{1,c}(t)\mathrm{d}t}.
\end{equation}


To calculate the folding time we need to solve the
first-passage probability $F_{1,c}(t)$
as a function of $t$ from the convolution relation (\ref{eq:FPTprobability}).
As an illustration, we first consider the case of $n=4$.
For the classical folding process,
we plot in Fig.~\ref{fig:collections}(e)
the $p^{(1)}_4 (t)$ and $p^{(4)}_4(t)$.
With these two time-dependent functions,
the first passage-probability $F_{1,4}(t)$ can be
further solved from the convolution relation (\ref{eq:FPTprobability})
by numerical iterations (see Fig.~\ref{fig:collections}(e)).
It is nonnegative and approaches to zero when $t$ goes to infinity.
This can be understood without difficulty
because the classical probability distribution
changes monotonously and approaches to its steady solution at the infinity time.
However, for a quantum walk the probability distribution oscillates in time.
We can see that the solved density matrix shown in Fig.~\ref{fig:collections}(a) and Fig.~S3  oscillates in time.
With this new characteristics in quantum walk, the value of the first-passage probability
solved directly from (\ref{eq:FPTprobability})
appears to be negative in certain time region (see Fig.~\ref{fig:collections}(b))
that is unphysical.


The zero point of $F_{1,4}(\tau^{}_0 )=0$ determines
the upper limit of the integration in the formula (\ref{eq:fdtime}).
In the simplest model with 4 residues,
the classical folding times $\tau^\mathrm{c}_\mathrm{fd}$  for the sequence subsets $Q_1$, $Q_2$ and $Q_3$ are
$6.0602$, $6.0351$ and $6.0180$ respectively.
Their corresponding quantum folding times are
$1.3208$, $1.2182$ and $0.9670$ respectively.
Clearly, the quantum folding is faster than the classical folding with about four to six times
even for the simplest model.
In the same way,
we calculate the quantum folding time for the forty-five situations
for the case with 6 residues
(see Tables SIII, SIV $\&$ SV in supplementary material).
One can see that the quantum folding is faster than the classical folding
with almost ten to hundred times or more.
The experimental observation~\cite{FoldingTimeExp} ever exhibited that the protein folding
is much faster than the theoretical prediction based on a random conformation search process.
To visualize more easily
we plot the quantum folding times $\tau^{}_\mathrm{fd}$
in Figs.~\ref{fig:collections} (g) to (i).
As a comparison,
we also plot the ratios of classical folding time $\tau^\mathrm{c}_\mathrm{fd}$
to the quantum folding time $\tau^{}_\mathrm{fd}$ on the same panels.
In those three histograms,
the longest folding time takes place for the sequence subsets $Q_{13}$, $Q_{38}$ and $Q_{42}$
while the shortest folding time occurs for the sequence subset $Q_{45}$, $Q_{29}$ and $Q_{29}$.
The largest ratios $\tau^\mathrm{c}_\mathrm{fd}/\tau^{}_\mathrm{fd}$ occur for the subsets $Q_{33}$, $Q_{31}$ and $Q_{41}$
but the smallest ratios occur for $Q_{17}$, $Q_{38}$ and $Q_{10}$.

In the above, we proposed a self-contained general theory to investigate protein folding problem quantum mechanically.
In terms of $H$-$P$ lattice model, one can always have a structure set $\mathscr{S}^{}_n$ for an amino-acid chain of any given number $n$ of residues.
With such a structure set, one can naturally define a connection graph $\mathscr{G}^{}_n$ by means of our definition of one-step folding.
Thus either a classical random walk or a quantum walk on the graph
can be solved with standard procedures.
The former implies a random conformational search
while the latter involves in fact a parallel search due to the quantum mechanical coherence~\cite{Christopher}.
The application of quantum walk has attracted more attentions~\cite{XuPeng} to study various contemporary topics
in recent years,
our present strategy may open a new avenue in the area of the application of quantum walks.
We have known if proteins were folded by sequentially sampling of all possible conformations,
the calculated folding time would be inevitably very large
because there is a very large number of degrees of freedom in an unfolded polypeptide chains.
We elucidated that the quantum evolution naturally helps us to understand a faster protein folding.
In terms of the concept of first-passage probability,
we can calculate the quantum protein folding time as the mean first-passage time.
It is worthwhile to mention that the first-passage probability solved from the conventional
convolution relation may take negative value in some time domain.
This is very important for the application of the quantum approach to an investigation of protein folding time.
According to our results for $n=4$ and $6$,
the quantum folding time is much shorter than that obtained from
classical random walk.
The presented theory is expected to bring in new insight
features of protein folding process.

The work is supported by National Key R \& D Program of China, Grant No. 2017YFA0304304,
and partially by the Fundamental Research Funds for the Central Universities.


\noindent{Supplementary material} is available in the online version of the paper.

\end{document}